\documentstyle[psfig]{mn}

\newif\ifAMStwofonts
\AMStwofontstrue

\ifoldfss
  \ifCUPmtlplainloaded \else
    \NewTextAlphabet{textbfit} {cmbxti10} {}
    \NewTextAlphabet{textbfss} {cmssbx10} {}
    \NewMathAlphabet{mathbfit} {cmbxti10} {} % for math mode
    \NewMathAlphabet{mathbfss} {cmssbx10} {} %  "   "    "
  \fi
  \ifAMStwofonts
    \ifCUPmtlplainloaded \else
      \NewSymbolFont{upmath} {eurm10}
      \NewSymbolFont{AMSa} {msam10}
      \NewMathSymbol{\upi}     {0}{upmath}{19}
      \NewMathSymbol{\umu}     {0}{upmath}{16}
      \NewMathSymbol{\upartial}{0}{upmath}{40}
      \NewMathSymbol{\leqslant}{3}{AMSa}{36}
      \NewMathSymbol{\geqslant}{3}{AMSa}{3E}

    \fi
  \fi
\fi % End of OFSS

\ifnfssone
  \newmathalphabet{\mathit}
  \addtoversion{normal}{\mathit}{cmr}{m}{it}
  \addtoversion{bold}{\mathit}{cmr}{bx}{it}
  \newmathalphabet{\mathbfit} % math mode version of \textbfit{..}
  \addtoversion{normal}{\mathbfit}{cmr}{bx}{it}
  \addtoversion{bold}{\mathbfit}{cmr}{bx}{it}
  \newmathalphabet{\mathbfss} % math mode version of \textbfss{..}
  \addtoversion{normal}{\mathbfss}{cmss}{bx}{n}
  \addtoversion{bold}{\mathbfss}{cmss}{bx}{n}
  \ifAMStwofonts
    \ifCUPmtlplainloaded \else
      \UseAMStwoboldmath
      \makeatletter
      \new@mathgroup\upmath@group
      \define@mathgroup\mv@normal\upmath@group{eur}{m}{n}
      \define@mathgroup\mv@bold\upmath@group{eur}{b}{n}
      \edef\UPM{\hexnumber\upmath@group}
      \new@mathgroup\amsa@group
      \define@mathgroup\mv@normal\amsa@group{msa}{m}{n}
      \define@mathgroup\mv@bold\amsa@group{msa}{m}{n}
      \edef\AMSa{\hexnumber\amsa@group}
      \makeatother
      \mathchardef\upi="0\UPM19
      \mathchardef\umu="0\UPM16
      \mathchardef\upartial="0\UPM40
      \mathchardef\leqslant="3\AMSa36
      \mathchardef\geqslant="3\AMSa3E
    \fi
  \fi
\fi % End of NFSS release 1

\ifnfsstwo
  \DeclareMathAlphabet{\mathbfit}{OT1}{cmr}{bx}{it}
  \SetMathAlphabet\mathbfit{bold}{OT1}{cmr}{bx}{it}
  \DeclareMathAlphabet{\mathbfss}{OT1}{cmss}{bx}{n}
  \SetMathAlphabet\mathbfss{bold}{OT1}{cmss}{bx}{n}
  \ifAMStwofonts
    \ifCUPmtlplainloaded \else
      \DeclareSymbolFont{UPM}{U}{eur}{m}{n}
      \SetSymbolFont{UPM}{bold}{U}{eur}{b}{n}
      \DeclareSymbolFont{AMSa}{U}{msa}{m}{n}
      \DeclareMathSymbol{\upi}{0}{UPM}{"19}
      \DeclareMathSymbol{\umu}{0}{UPM}{"16}
      \DeclareMathSymbol{\upartial}{0}{UPM}{"40}
      \DeclareMathSymbol{\leqslant}{3}{AMSa}{"36}
      \DeclareMathSymbol{\geqslant}{3}{AMSa}{"3E}
    \fi
  \fi
\fi % End of NFSS release 2

\ifCUPmtlplainloaded \else
  \ifAMStwofonts \else % If no AMS fonts
    \def\upi{\pi}
    \def\umu{\mu}
    \def\upartial{\partial}
  \fi
\fi

\makeatletter
\newcounter{cureqno}
    {\let\theequation\@curtheeqn%
    \setcounter{equation}{\value{cureqno}}}
\makeatother

\def\figsize{\ifSFB@referee\epsfxsize=0.5\hsize\else\epsfxsize=\hsize\fi}
\newif\ifdraft \def\draft{\ifSFB@referee\drafttrue\fi}
\makeatother
\def\eq#1#2 {\begin{equation} \if!#2!{#1}\else\label{#1}#2\fi \end{equation}
      \ifdraft\if!#2!\else\marginpar{\small #1}\fi\fi}
\def\eqarray#1#2 {\begin{eqnarray}\if!#2!{#1}\else\label{#1}#2\fi\end{eqnarray}
      \ifdraft\if!#2!\else\marginpar{\small #1}\fi\fi}

\def\Ref{\bibitem{}}

\def\mic  {\hbox{$\umu$m}}

\def\symbol#1{\hbox{$#1$}}

\def\about{\symbol{\sim}}

\def\La   {\symbol{L_{\rm AGB}}}
\def\Lo   {\symbol{L_\odot}}

\def\Fa   {\symbol{F_{12}}}
\def\Fb   {\symbol{F_{25}}}
\def\Fbol {\symbol{F_{\rm bol}}}
\def\Lbol {\symbol{L_{\rm bol}}}
\def\Mdot {\symbol{\dot{M}}}
\def\Lo     {\hbox{L$_\odot$}}
\def\Mo     {\hbox{M$_\odot$}}
\def\Ro     {\hbox{R$_\odot$}}
\def\c12  {\hbox{[25]-[12]}}
\def\c23  {\hbox{[60]-[25]}}
\def\R    {\symbol{R}}

\pagerange{\pageref{firstpage}--\pageref{lastpage}}
\def\Draft{Submitted to MNRAS}

\pubyear{2002}

\title[The Galactic Distribution of AGB Stars]
      {The Galactic Distribution of Asymptotic Giant Branch Stars}

\author[Jackson, Ivezi\'{c} and Knapp]
     {Tom Jackson, \v{Z}eljko Ivezi\'{c} and G.R. Knapp \\
      Princeton University, Department of Astrophysical Sciences, \\
      Princeton, NJ 08544--1001; jackson, ivezic, gk@astro.Princeton.edu}

\date{\Draft}

\begin{document}

\maketitle

\label{firstpage}

\begin{abstract}
We study the Galactic distribution of \about 10,000 Asymptotic
Giant Branch (AGB) stars selected by IRAS colors and variability
index. The distance to each star is estimated by assuming a narrow
luminosity function and a model-derived bolometric correction.
The characteristic AGB star luminosity, $L_{AGB}$, is determined
from the condition that the highest number density must coincide with
the Galactic bulge. Assuming a bulge distance of 8 kpc, we determine
$L_{AGB}$\about 3,500 \Lo, in close agreement with values obtained
for nearby AGB stars using the HIPPARCOS data.

We find that there are no statistically significant differences in the
Galactic distribution of AGB stars with different IRAS colors, implying
a universal density distribution. The direct determination of this
distribution shows that it is separable in the radial, $R$, and vertical,
$z$, directions. Perpendicular to the Galactic plane, the number density
of AGB stars is well described by an exponential function with a vertical
scale height of 300 pc. In the radial direction the number density of AGB
stars is constant up to $R \sim$ 5 kpc, and then it decreases exponentially
with a scale length of \about 1.6 kpc. This fall-off extends to at least
12 kpc, where the sample becomes too small. The overall normalization 
implies that there are about 200,000 AGB stars in the Galaxy.

We estimate the [25]-[12] color distribution of AGB stars for an unbiased
volume-limited sample. By using a model-dependent transformation between the
color and mass-loss rate, \Mdot, we constrain the time dependence of \Mdot.
The results suggest that for 10$^{-6}$ \Mo yr$^{-1} < \Mdot < 10^{-5}$ \Mo
yr$^{-1}$ the mass-loss rate increases exponentially with time. We find only
marginal evidence that the mass-loss rate increases with stellar mass.
\end{abstract}

\section{                         Introduction                        }

Studies of the stellar distribution within the Galaxy can provide
information on its formation mechanism(s) and subsequent evolution.
While a significant amount of data about the Galaxy has been collected over
the years, the knowledge about the stellar distribution in the Galactic
plane is still limited to a few kpc from the Sun (Mihalas \& Binney 1981,
Binney \& Tremaine 1987) by interstellar dust extinction because the
visual  extinction is already 1 mag at a distance of only about
0.6 kpc (Spitzer, 1978).

The interstellar dust extinction decreases with wavelength and is all but
negligible beyond about 10 \mic, even for stars at the Galactic center.
For this reason, the analysis of the data obtained by the Infrared Astronomical
Satellite (IRAS, Beichman {\em et al.} 1985) has significantly enhanced our
knowledge of the stellar distribution in the Galactic disk and bulge.
IRAS surveyed 96\% of the sky at 12, 25, 60 and 100 \mic, with the
resulting point source catalog (IRAS PSC) containing over 250,000 sources.
The colors based on IRAS fluxes\footnote{Except when discussing bolometric
flux, the implied meaning of ``flux" is the flux density.} can efficiently be
used to distinguish pre-main sequence from post-main sequence stars, and to
study characteristics of their dust emission (e.g. van der Veen \& Habing 1987,
Ivezi\'c \& Elitzur 2000, hereafter IE00).

Soon after the IRAS data became available it was realized that properly
color-selected point sources clearly outline the disk and the bulge
(Habing et al. 1985). The color selection corresponds to OH/IR stars,
asymptotic giant branch (AGB) stars with very thick dust shells due
to intensive mass loss (for a detailed review see Habing 1996). Because
these stars have large luminosity ($\about 10^3-10^4$ \Lo), and because
a large fraction of that luminosity is radiated in the mid-IR due to
reprocessing by surrounding dust, they are brighter than the IRAS faint
cutoff ($<$ 1 Jy at 12 \mic) even at the distance of the Galactic center.
The availability of IRAS data soon prompted several detailed studies of the
Galactic distribution of AGB stars. Common features in all these studies are
the sample selection based on the IRAS \Fa\ vs. [25]-[12] color-magnitude
diagram, and the determination of the bolometric flux by utilizing a model
based bolometric correction.

Habing et al. (1985) discuss \about7,000 bulge stars selected by
0.5 $<$ \Fa/\Fb $<$ 1.5 and 1 Jy $<$ \Fa $<$ 5 Jy, in two areas defined by
$|b| < 10^\circ, |l| < 10^\circ$, and $|b| < 10^\circ, 10^\circ < l < 30^\circ$.
Assuming that for all stars \Fa = 2.3 Jy, and a bolometric correction calculated
for a T=350 K black body, they derive \Lbol = 2600 \Lo (for the Galactic center
distance of 8 kpc, Reid 1989). They also find that 25\% of the selected sources have IRAS
variability index (the probability that a source is variable, expressed in
percent, for definition see Section 2.2) larger than 99, as compared
to 13\% for all stars from the IRAS PSC. This difference is in good
agreement with the known long period variability of AGB stars. Habing et al.
(1985) also noted that despite this agreement, their sample may still
be severely contaminated by planetary nebulae. They classified 2\% of
stars from the sample using supplemental data and found the same fractions of
variable stars and planetary nebulae.

Rowan-Robinson \& Chester (1987) select bulge stars by requiring \Fa $>$ 1 Jy
(interpreted as the IRAS confusion limit) and $|b| < 10^\circ, |l| < 10^\circ$.
They assumed that all these objects {\it in the direction} of the Galactic
center are actually {\it at} the Galactic center. This assumption allowed
them to determine the median luminosity (\about 3,000 \Lo) and to place
an upper limit on the width of the luminosity function, which was found
to be very narrow (the root-mean-square scatter is about a factor of 2).
They also determined the distribution of colors and transformed it
into a distribution of the shell optical depth by using model-dependent
transformations. Assuming that this luminosity function and optical
depth distribution also apply to the Galactic disk, and that the disk volume
density distribution can be parametrized as two separable exponential functions in
the vertical, $z$, and radial, \R, directions, they derive a scale
height of 250 pc and a scale length of 6 kpc.

Habing (1988) extended the analysis to disk sources in several areas on the
sky defined as thin strips parallel to the Galactic equator, with a total area
of 1200 deg$^2$, or about 3\% of the sky. AGB stars are selected by requiring
$\Fa >$ 1 Jy, 0.3 $<$ \Fb/\Fa $<$ 3.8 and q$_{12}$ = q$_{25}$ = 3,
where q$_{12}$ and q$_{25}$ are IRAS flux qualities at 12 and 25 \mic\
(1=upper limit, 2=low quality, 3=high quality). Habing does not include the
60 \mic\ flux in the source selection because the resulting number of sources
is too small, but does exclude those sources for which F$_{60}>$\Fb\ is reliably
measured. This sample was used to constrain the luminosity
function and spatial distribution of stars by fitting number counts in the
selected areas. The spatial distribution is assumed to be separable: a
sech$^2$ function for the $z$ direction with an \R-independent scale
height (this was motivated by results obtained for other galaxies),
and an exponential for the $R$ direction. Habing finds that
the models are not unique, and it is hard to find a best-fitting one.
He concludes that the sample contained two populations with either a similar
spatial density and different luminosities, or a similar luminosity but different
spatial distributions. Habing prefers the second option and argues that the
results present evidence for the thick disk proposed by
Gilmore and Reid (1983), and for a thin disk cutoff at \R \about 10 kpc.
He also points out that the luminosity function in the disk is similar
to that in the bulge, providing support for the earlier assumption by
Rowan-Robinson and Chester (1987).

Blommaert, van der Veen \& Habing (1993) extended the study by Habing (1988)
with the aim of determining the sample contamination. They obtained near-IR
photometry and OH maser measurements for 53 sources which are located outside
the solar circle and have \Fb $>$ \Fa\ (region IIIb of the IRAS color-color
diagram, as defined by van der Veen and Habing, 1988). Although this subsample
is expected to have the least amount of contamination by non-AGB
stars, they find that \about 55\% of objects are not AGB stars.  The contaminating
sources have overestimated distances due to both underestimated bolometric corrections
and overestimated luminosities, and resulted in spurious evidence for a thick disk
and the thin disk cutoff.

These pioneering IRAS-based studies suggested that the luminosity function
for AGB stars is rather narrow and centered around $L$ \about 3,000 \Lo,
and appears not to vary strongly with position in the Galaxy.
The IRAS number counts can be reasonably well fitted by assuming a spatial
distribution of AGB stars which is separable in $z$ and $R$, and described by
exponential functions with a scale height of 250 pc and a scale length
of 6 kpc, respectively.

In this work we revisit the problem of constraining the Galactic
distribution of AGB stars using IRAS data. There are several factors
motivating us to perform a study similar to those listed above:

\begin{enumerate}
\item
The study by Blommaert, van der Veen \& Habing (1993) showed that the
variability index is a reliable indicator of AGB stars, while
samples selected by colors alone can be significantly contaminated.
Thus, it seems prudent to complement the sample color selection by using
the IRAS variability index and repeat the analysis. Further,
the understanding of the IRAS color-color diagrams and the distribution
of various types of dusty star has advanced since the early studies of
IRAS data, and can be used to select cleaner, more reliable samples
(e.g. van der Veen \& Habing 1988, IE00).
\item
The spectral energy distribution (SED) models for AGB stars and their
dependence on the various stellar parameters are also better understood.
In particular, more reliable bolometric corrections are available, and
the IRAS colors are recognized as an indicator of the mass-loss rate
(Habing 1996, and references therein).
\item
The HIPPARCOS data allowed direct determination of the bolometric
luminosity for AGB stars in the solar neighborhood. The luminosity
distribution of nearby AGB stars is very narrow and centered around
$L$ \about 3,000 \Lo\ (Knauer, Ivezi\'{c} \& Knapp 2001). Since it is
very similar to the luminosity function obtained for the much redder bulge
stars, it appears that the luminosity function is similar throughout the
Galaxy and not very dependent on stellar color. Due to this similarity,
it is possible to postulate a universal narrow luminosity function
and estimate the distance to each star irrespective of its color
and position in the Galaxy. This allows a {\it direct} study of their galactic
distribution, rather than constraining it indirectly by modeling the number
counts vs. flux relation.
\item
All previous studies {\em assumed} that the number density distribution
is separable in $z$ and $R$. Although the available IRAS data
appear sufficient to explicitly test this assumption, this has not
yet been done. In addition, subsamples of AGB stars with different
colors (reflecting different mass-loss rates) can be formed, allowing
a study of differences in their Galactic distribution.
\end{enumerate}

In Section 2 we discuss the IRAS data and the method developed for
selecting AGB stars. The determination of the Galactic distribution
of \about 10,000 selected stars is described in Section 3, and in
Section 4 we analyze the time evolution of AGB mass loss.
The results are summarized and discussed in 5.

\section{                    Selection Method                     }

\subsection{                 The IRAS PSC Data                 }

%%%%%%%%%%%%%%%%%%%%%%%%%%%%%%%%%%%%%%%%%%%%%%%%%%%%%%%%%%%%%%%%%%%%%%%%%%%%
\begin{figure}
%\begin{minipage}{\textwidth}
\centering \leavevmode \psfig{file=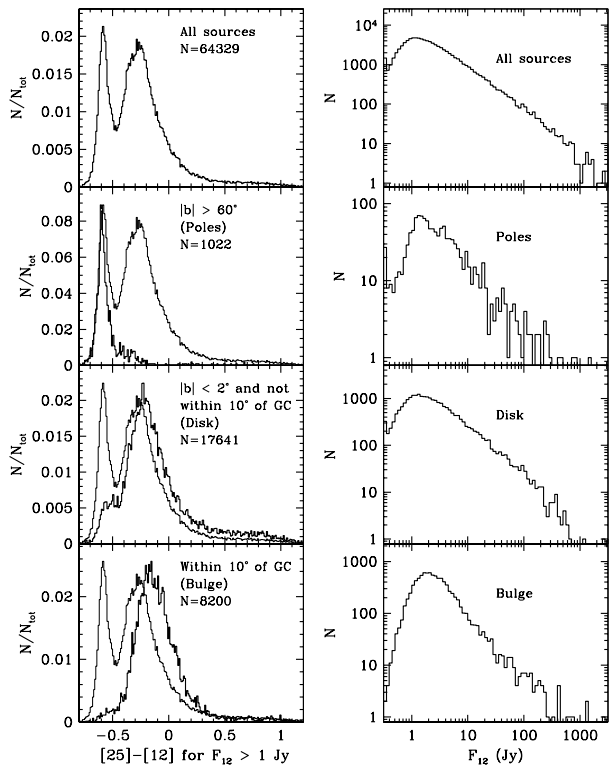,width=\hsize,clip=}
\caption{The top two panels show IRAS \Fa\ flux distributions for the 88,619
sources with both 12 \mic\ and 25 \mic\ flux qualities greater than 1 (right),
and [25]-[12] color distributions for the subset of 64,329 sources with
\Fa $>$ 1 Jy (left). The following three rows show analogous histograms for
subsamples selected by their galactic coordinates, as marked in the panels.
Their color distribution is shown by the thick line, and compared to the color
distribution of the whole sample shown by the thin line.
\label{IRAS22xx}
}
%\end{minipage}
\end{figure}
%%%%%%%%%%%%%%%%%%%%%%%%%%%%%%%%%%%%%%%%%%%%%%%%%%%%%%%%%%%%%%%%%%%%%%%%%%%%

The IRAS color-color and color-magnitude diagrams have been
extensively used to select and classify various types of
dusty star (e.g. van der Veen and Habing 1988, Walker et al.
1989). Most recently, IE00 showed that all IRAS sources can be separated
into 4 classes, including  AGB stars. They obtain a very clean
classification by using a scheme which requires at least 3 IRAS fluxes.
For AGB stars the three best fluxes typically are 12 \mic,
25 \mic, and 60 \mic\ fluxes. However, as already pointed out by
Habing (1988), the requirement that the IRAS flux at 60 \mic\ be of
good quality significantly decreases (by more than a factor 4)
the number of available sources. Further, even for stars
with formally good 60 \mic\ flux, there is significant contamination
by interstellar cirrus emission (Ivezi\'c \& Elitzur 1995, hereafter
IE95). For this reason, we are forced to use only the IRAS fluxes at
12 \mic\ and 25 \mic, and the resulting color [25]-[12] = log(\Fb/\Fa).

There are 88,619 sources in the IRAS PSC with both the 12 \mic\ and 25 \mic\
flux qualities greater than 1. Their [25]-[12] color and \Fa\ flux
distributions are shown in the top two panels in Figure \ref{IRAS22xx}.
The right panel indicates that the sample is complete to
\Fa \about 1 Jy. The left panel shows the color distribution
for 64,329 sources with \Fa $>$ 1 Jy. There are two obvious peaks.
The peak at [25]-[12] = -0.6 represents stars without dust
emission, and the peak at [25]-[12] = -0.25 is dominated by
AGB stars (IE00).

The color and flux distributions of the stars in the sample
are very dependent on the galactic coordinates. The two panels
in the second row in Figure \ref{IRAS22xx} show the color and flux
distributions for a subsample of 1,022 stars towards the galactic
poles ($|b| > 60^\circ$) compared to the color distribution of the whole sample.
As evident, the sample is dominated by dust-free
stars. On the contrary, the disk stars and stars towards the bulge
are dominated by AGB stars, as evident in the panels in the
last two rows. The strong dependence of the color and flux
distributions on the galactic coordinates indicates the rich
information on the Galactic structure encoded in these data.

\subsection{    The AGB Sample Selection Criteria       }

%%%%%%%%%%%%%%%%%%%%%%%%%%%%%%%%%%%%%%%%%%%%%%%%%%%%%%%%%%%%%%%%%%%%%%%
\begin{figure}
%\begin{minipage}{\textwidth}
\centering \leavevmode \psfig{file=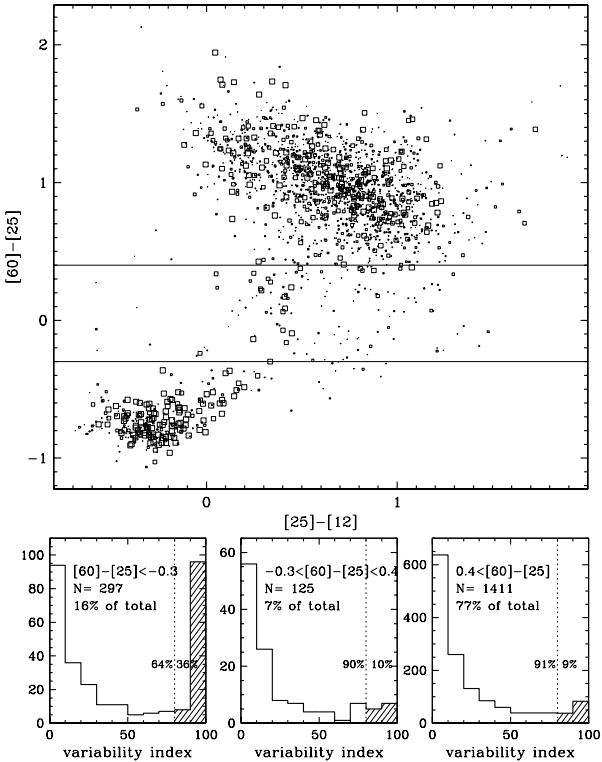,width=\hsize,clip=}
\caption{The top panel shows IRAS [60]-[25] vs. [25]-[12] color-color
diagram for 1,624 sources with both the 12 \mic, 25 \mic, and 60 \mic\
flux qualities equal to 3. The symbol size is proportional to the
variability index. The bottom panels show the variability index
distribution for three subsamples selected by [60]-[25] color.
\label{IRclasses}}
%\end{minipage}
\end{figure}
%%%%%%%%%%%%%%%%%%%%%%%%%%%%%%%%%%%%%%%%%%%%%%%%%%%%%%%%%%%%%%%%%%%%%%%

Figure 2 in IE00 shows that sources with [25]-[12] $<$ -0.2 are
dominated by AGB stars, and that their fraction becomes negligible
for [25]-[12] $>$ 0.2, where the sample is dominated by young stellar
objects. An optimal separation line between the AGB
stars and other sources (mostly young stellar objects and planetary
nebulae) is [25]-[12]=0. While it is tempting to define the sample
by using only this criterion, Blommaert, van der Veen \& Habing
(1993) showed that color-selected samples may be significantly
contaminated by sources other than AGB stars such as pre-main sequence stars
and planetary nebulae. They also pointed out
that the IRAS variability index, $var$, is a reliable indicator of
AGB stars (which are known to vary on time scales of \about 1 year).
This finding was further reinforced by Allen, Kleinmann \& Weinberg
(1993) who found that the IRAS stars with high variability index are
dominated by AGB stars.

The IRAS variability index was estimated by comparing the number of
sources with correlated flux excursions exceeding $m\sigma$ at 12 and
25 \mic, $N_{c}(m)$, with the number of sources showing anti-correlated
flux excursions exceeding $m\sigma$, $N_{a}(m)$, where $\sigma$ is the
measurement error. The probability that a source is variable is computed
from
\eq{
           p = (a-b) / (a+b),
}
where $a = N_{c}(m)/N_{c}(0)$, and $b = N_{a}(m)/N_{a}(0)$ (IRAS Explanatory
Supplement\footnote{The IRAS Explanatory Supplement is available at
http://space.gsfc.nasa.gov/astro/iras/docs/exp.sup}, equation V.H.3).
In order to determine an optimal AGB selection cut for the variability
index, we investigated the distribution of the variability index,
$var = p \times 100\%$,
in the IRAS 12-25-60 color-color diagram. Figure \ref{IRclasses} shows the
[60]-[25] vs. [25]-[12] color-color diagram for the 1,624 sources with the highest
quality fluxes at 12 \mic, 25 \mic\ and 60 \mic.  These
sources have three reliable fluxes, and thus can be reliably classified
by their position in the [60]-[25] vs. [25]-[12] color-color diagram,
as described by IE00. The two horizontal lines divide the diagram into
regions dominated by AGB stars ([60]-[25] $<$ -0.3), planetary nebulae
(-0.3 $<$ [60]-[25] $<$ 0.4) and young stellar objects ([60]-[25] $>$ 0.4).
The number of sources in each region is listed in the first row of Table 1.
Sources with high variability index ($var >$ 80) are found throughout
the diagram, but at a higher rate among AGB stars. This is better seen
in the three histograms shown at the bottom of the figure, where only
AGB stars show a strong excess of high variability index, $var > 80$.
There are 36\% of
AGB stars with $var >$ 80, while only 10\% or less of other sources show
such a high variability index.

\begin{figure}
\centering \leavevmode \psfig{file=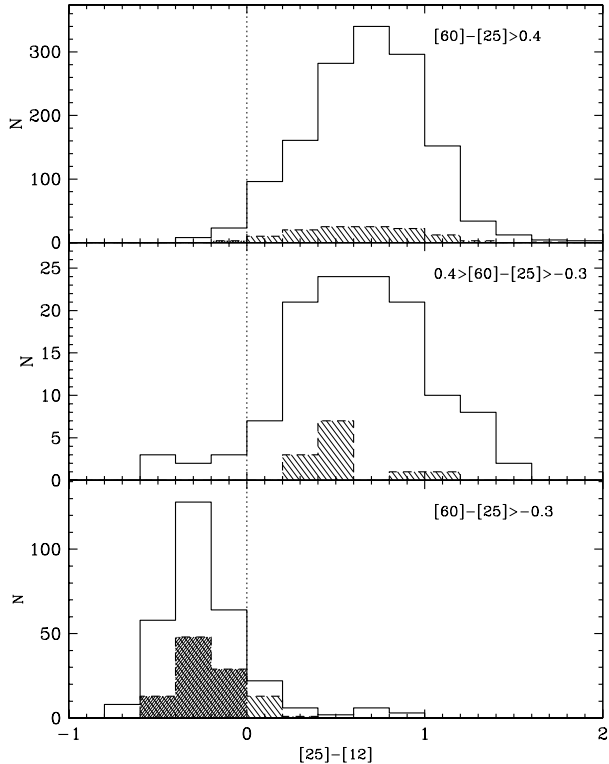,width=\hsize,clip=}
\caption{The comparison of the selection methods in three bins of
[60]-[25] color. The solid line shows the (unshaded) histogram of all
sources in a given [60]-[25] bin, and the dashed line shows the (hatched)
histogram for sources with the variability index greater than 80.
The cross-hatched histogram indicates sources with both high variability
index and [25]-[12] $<$ 0.\label{selection}}
\end{figure}

We adopt $var >$ 80, which is the minimum of the $var$ distribution in
the whole sample, as the additional selection criterion for AGB stars.
According to the IRAS Explanatory Supplement (Section VII.D.3), this
variability cut roughly corresponds to a variability amplitude of about
0.2 mag. The sample of 1,624 sources with high-quality fluxes can be used
to estimate the selection efficiency. Figure \ref{selection} compares the
selection method for three types of source separated by their [60]-[25] color:
all sources, sources with $var >$ 80, and with high variability index and 
[25]-[12] $<$ 0. The number of sources in each category is listed in Table 1. 
There are only \about 3\% of non-AGB stars\footnote{Following IE00, we assume 
that all stars with [25]-[12] $<$ 0 and [60]-[25] $< -0.3$ are AGB stars.} 
in the sample selected by using both the color and variability criteria. 
This is a significant improvement compared to the sample selected only by 
color which contains \about 12\% of non-AGB stars.

%%%%%%%%%%%%%%%%%%%%%%%%%%%%% Table 1 %%%%%%%%%%%%%%%%%%%%%%%%%%%%%
\begin{table}
\begin{center}

\begin{tabular}{rrrrr}
\hline
[60]-[25]:  & $<$ -0.3 & -0.3 -- 0.4 & $>$ 0.4 & All \\
\hline \hline
N(total)       & 297  &  119   &       1208   &  1624  \\
N([25]-[12]$<$0)  & 258  &    7   &         28   &   293  \\
N($var  > 80$) & 104  &   12   &        110   &   226  \\
N(selected)    &  90  &    0   &          3   &    93  \\
\hline
\end{tabular}
\caption{The statistics for color and variability selection cuts (see text).}
\smallskip
\end{center}
\end{table}
%%%%%%%%%%%%%%%%%%%%%%%%%%%%%%%%%%%%%%%%%%%%%%%%%%%%%%%%%%%%%%%%%%%%%%%%%%%%%%

%%%%%%%%%%%%%%%%%%%%%%%%%%%%%%%%%%%%%%%%%%%%%%%%%%%%%%%%%%%%%%%%%%%%%%%
%%%%%%%%%%%%%%%%%%%%%%%%%%%%%%%%%%%%%%%%%%%%%%%%%%%%%%%%%%%%%%%%%%%%%%%

The adopted variability cut selects 35\% of all AGB stars with
[25]-[12] $<$ 0. While using only the color selection would thus increase
the sample size by almost a factor of three, the analysis by Habing (1988)
and Blommaert, van der Veen \& Habing (1993) showed that the sample
contamination may significantly affect the derived conclusions.
That is, although the variability cut significantly decreases the sample size,
it also decreases the (fractional) contamination by non-AGB sources by
a factor of 4.

The variability cut, while efficient in excluding non-AGB sources, may
introduce a selection bias. For example, the variability detection could
be significantly dependent on the color, flux or position of a source.
Indeed, only about 70\% of the sky was surveyed three times during
the IRAS mission, while 20\% was observed only twice, and thus
the variable sources in some parts of the sky were more likely
to be detected than others. Fortunately for the analysis presented in
this paper, the IRAS scans were arranged along lines of constant ecliptic
longitude, and consequently this effect is not strongly correlated with
galactic structure (we further discuss it in Section 3.5)

%%%%%%%%%%%%%%%%%%%%%%%%%%%%%%%%%%%%%%%%%%%%%%%%%%%%%%%%%%%%%%%%%%%%%%%
\begin{figure}
\centering \leavevmode \psfig{file=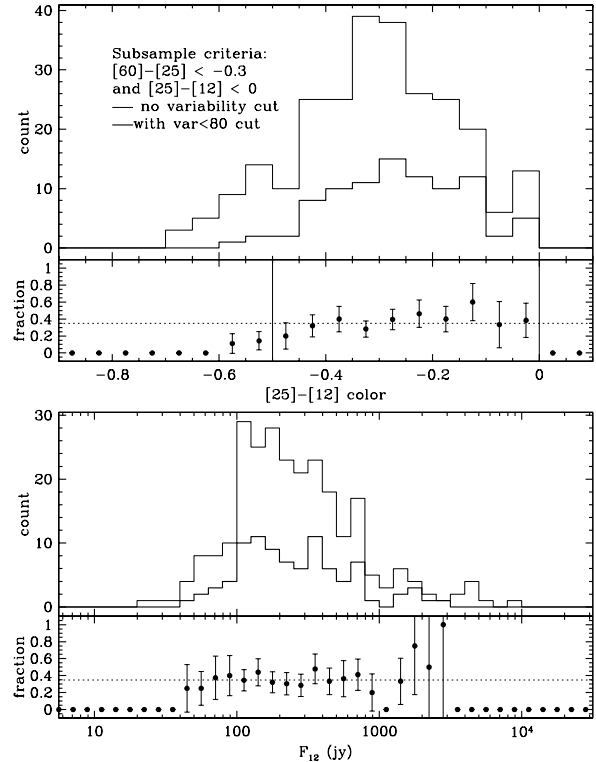,width=\hsize,clip=}
\caption{The top panel shows the [25]-[12] color histogram for
the whole sample by the thin line, and for sources with $var > $ 80 by
the thick line. The symbols show the fraction of the latter in the whole
sample for each color bin. The bottom panel shows the analogous
histograms and the corresponding fraction when the sample is binned by
\Fa\ flux.\label{bias}}
\end{figure}
%%%%%%%%%%%%%%%%%%%%%%%%%%%%%%%%%%%%%%%%%%%%%%%%%%%%%%%%%%%%%%%%%%%%%%%

We analyze the possibility of a selection bias with respect to flux
and color by comparing the distributions of the \Fa\ flux and [25]-[12] color
for the sample of 293 stars with high quality fluxes and classified as
AGB stars, to the distributions for a subsample of 93 stars with $var>80$.
The top panel in Figure \ref{bias} compares the [25]-[12] color histogram for
the whole sample and for sources with $var > $ 80. As evident, there is no strong 
dependence of this
fraction on color in the -0.5 $<$ [25]-[12] $<$ 0 range. The bottom panel in
Figure \ref{bias} shows the analogous histograms and the corresponding
fraction when the sample is binned by \Fa\ flux. Again, there is no
significant correlation between the fraction of selected sources and
the \Fa\ flux. It should be noted that the faint limit of this sample
is brighter than for the sample shown in Figure \ref{IRAS22xx}
(40 Jy vs 1 Jy) due to the difference in required flux qualities.
Thus, the possibility of selection biases for faint sources cannot be
fully excluded.

In summary, we require that the candidate AGB stars have flux qualities
at 12 \mic\ and 25 \mic\ greater than 1, [25]-[12] $<$ 0, and variability
index greater than 80. These selection criteria result in a sample of
10,240 stars, with the sample completeness of 35\%, and contamination by
non-AGB stars of about 3\%.

With only two IRAS fluxes it is impossible to reliably distinguish stars
with silicate dust from stars with carbon dust. It is estimated that about
10\% of AGB stars observed by IRAS have carbon dust (e.g. Wainscoat {\em
et al.} 1992). We adopt this estimate in the remainder of our analysis,
and note that it could be incorrect by as much as a factor of 2.
Nevertheless, we found that varying the assumed fraction from 5\% to 20\%
has only a minor effect on our results.

\section{           The Galactic Distribution of AGB Stars            }

Most previous studies utilized number counts as a tool to infer the
galactic distribution of AGB stars (see Section 1). For chosen descriptions
of the stellar density distribution and luminosity function, whether in analytic
or non-parametric forms, the models are constrained by fitting
the predicted counts vs. flux relations to the observed counts for
different regions on the sky. That is, the galactic distribution
is constrained only {\it indirectly} through its effect on the observed
source counts. Here we follow a different approach: we estimate
the distance to each star in the sample and {\it directly} determine
their galactic distribution.

We estimate distances to individual stars by exploiting the observation
that the bolometric luminosity function for AGB stars is fairly narrow,
and approximate it by assuming that all stars have the same characteristic
luminosity, \La. Then the distance, $D$, to a star with bolometric flux,
$F_{bol}$, is estimated from $4 \pi D^2 = \La/F_{bol}$. We utilize the stellar
angular distribution towards the Galactic center and the known distance to the
Galactic center, and determine the value of the characteristic luminosity, \La.
By assuming circular symmetry of the Galactic bulge and the disk, we confirm
{\it a posteriori} that the luminosity function is very narrow.

\subsection{  The IR Bolometric Correction for AGB stars}

The approach followed here, as well as in most other studies, depends on
a bolometric correction to determine the bolometric (total) flux of star
as a function of its measured IRAS \Fa\ and \Fb\ fluxes. It is a standard
procedure in the optical wavelength range to use the flux and color of a
star to determine its bolometric flux via
\eq{
          m_{bol} = m_1 + BC(m_2-m_1),
}
where $BC$ is the bolometric correction, $m_1$ and $m_2$ are magnitudes at two
different wavelengths, and $m_{bol}$ is the bolometric magnitude (e.g. Allen 1973).
It is possible to determine the bolometric flux by using only two measurements
because stellar SEDs are by and large a function of a single parameter: the effective
temperature. While the gravity and metallicity also play a role, their
influence on the broad-band fluxes is typically minor ($\la$ 0.1-0.2 mag.,
e.g. Lenz {\em et al.} 1998).

It is not {\it a priori} clear that an analogous procedure can be
used for AGB stars in the IR range. AGB stars have very red SEDs
because their stellar radiation is absorbed by dusty circumstellar
envelope and reradiated at IR wavelengths. The SED models for
AGB stars typically involve many input parameters (stellar temperature,
mass and luminosity, mass-loss rate, dust properties, geometrical
dimensions) and it seems that most of them can significantly affect
the SED. Nevertheless, it was established empirically that it is possible
to construct a well-defined IR bolometric correction for AGB stars
(Herman, Burger \& Penninx 1986, van der Veen \& Rugers 1989). For stars
with good photometric wavelength coverage the bolometric flux can be determined
by direct integration, and a good correlation is found between the ratio
$F_{bol}$/\Fa\ and the [25]-[12] color, such that
\eq{
               F_{bol} = \Fa * BC([25]-[12]),
}
where BC([25]-[12]) is the ``infrared" bolometric correction.

The existence of a reasonable IR bolometric correction is understood as
a consequence of the scaling properties of the radiative transfer
equation, and the universality of the dust density distribution in
envelopes around AGB stars (Rowan-Robinson 1980, IE95, Ivezi\'{c} \& Elitzur
1997, hereafter IE97, Elitzur \& Ivezi\'{c} 2001, herafter EI). While individual
parameters such as e.g.
luminosity and mass-loss rate, affect the SED, the SED for given dust grains is
fully parameterized by a single parameter, overall optical depth at some
fiducial wavelength. Since all dimensionless quantities derived from the
SED are functions of the optical depth, including the ratio $F_{bol}/\Fa$
and [25]-[12] color, the ratio $F_{bol}/\Fa$ then must be a function of
the [25]-[12] color. That is, while the effective temperature by and large
controls the SEDs of dust-free stars, the SEDs of dust-enshrouded stars
are essentially fully controlled by the dust optical depth\footnote{For
optical depths so small that dust emission is negligible, the bolometric
correction becomes the bolometric correction of a naked star, which is similar
for all such AGB stars because they span a very narrow range of effective
temperature. These stars have [25]-[12] $<-0.5$ and are not included in 
the final sample discussed here.}.

%%%%%%%%%%%%%%%%%%%%%%%%%%%%%%%%%%%%%%%%%%%%%%%%%%%%%%%%%%%%%%%%%%%%%%%
\begin{figure}
\centering \leavevmode \psfig{file=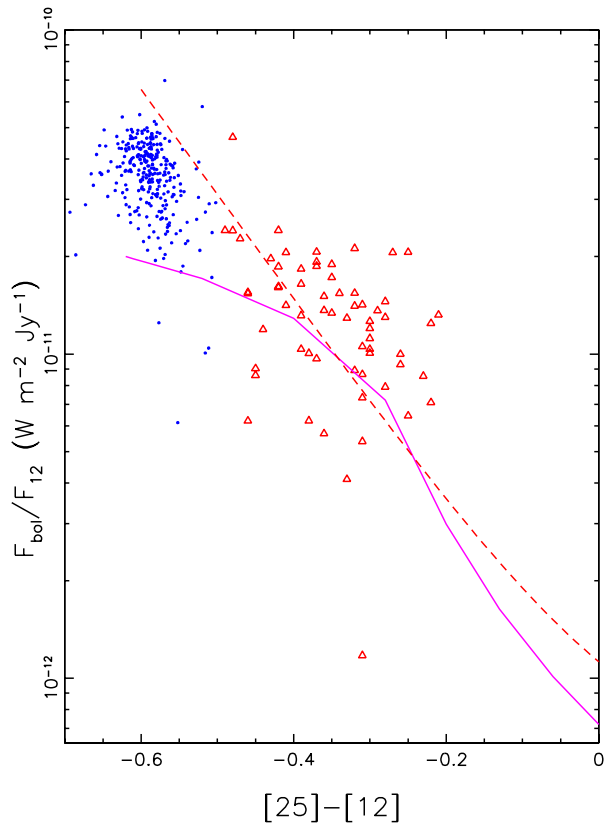,width=\hsize,clip=}
\caption{The ``infrared" bolometric correction for AGB stars. Symbols
show data for dusty (triangles) and dust-free (dots) stars. The solid
line is a model-derived bolometric correction used in this work, and 
the dashed line is a fit used by van der Veen \& Breukers (1989). 
\label{BC}}
\end{figure}
%%%%%%%%%%%%%%%%%%%%%%%%%%%%%%%%%%%%%%%%%%%%%%%%%%%%%%%%%%%%%%%%%%%%%%%

We utilize a bolometric correction derived from the models described
in IE95 and computed by the DUSTY code (Ivezi\'{c}, Nenkova \& Elitzur
1997). In particular, we use ``warm'' silicate grains from Ossenkopf, Henning 
\& Mathis (1992), assume that stellar spectrum is a 3000 K black body, and 
that the highest dust temperature is 500 K. These values provide best 
agreement with the data discussed by Knauer, Ivezi\'{c} \& Knapp (2001),
and shown as symbols in Figure \ref{BC}. For comparison, we show the
bolometric correction determined by van der Veen \& Breukers (1989), for
a different sample of AGB stars, as the dashed line. 

For negligible optical depths ([25]-[12] \about -0.6) the bolometric correction
has the value corresponding to the input stellar spectrum. This value varies
with the stellar temperature as T$^{-3}$ because the IRAS wavelengths are
in the Rayleigh-Jeans domain. As the optical depth increases, the SED is
shifted towards longer wavelengths, the ratio of the \Fa\ flux and bolometric
flux increases (i.e. BC decreases), and the [25]-[12] color becomes redder.
Note that for a given \Fa\, the corresponding bolometric flux decreases
as the [25]-[12] color becomes redder. The best-fit value for the highest dust 
temperature is about 200 K lower than usually assumed when modeling the SEDs
of AGB stars. However, this low value is in agreement with Marengo, Ivezi\'c 
\& Knapp (2001) who show that AGB stars (with silicate dust) showing semi-regular 
variability require models with somewhat lower dust temperatures (\about 300 K) 
than stars exhibiting Mira-type variability (\about 750 K). 

We conclude that it is possible to use IR bolometric correction for AGB stars
to estimate their bolometric fluxes from IRAS \Fa\ and \Fb\ fluxes, albeit
with an uncertainty up to a factor of 2. We show below that
the actual uncertainty in derived bolometric fluxes seems not larger than
\about 50\%.

\subsection{             Determination of \La           }

By assuming that all AGB stars have the same luminosity, \La, and adopting
a model-derived bolometric correction, the distance to each star is estimated
from
\eq{
\label{dist}
    D = \left({\La \over 4 \pi \Fa \, BC([25]-[12])}\right)^{1/2}.
}
The characteristic luminosity, \La, is, of course, unconstrained in
the case of an individual star without an independent distance estimate.
Nevertheless, \La\ can be estimated for an {\em ensemble} of stars because
the stellar distribution has a local maximum at the Galactic center, and the
distance to the Galactic center is well determined (we use 8 kpc, Reid 1989).

As shown by Habing {\em et al.} (1985), ``properly" color-selected IRAS
point sources clearly outline the Galactic disk and the bulge. This observation
unambiguously indicates that IRAS observed AGB stars as far as the Galactic
bulge; however, it is not clear what is the limiting distance to which 
IRAS detected AGB stars. The incompleteness effects close to the faint 
sensitivity limit may bias the estimate of \La.

We examined this limiting distance by a method that does not depend
significantly on the sample completeness near the faint cutoff.
We compare the counts ratio of stars seen directly towards
the bulge, and disk stars (both selected by applying angular masks, see below)
selected from narrow color and flux bins. Since both subsamples have the same observed
flux, they are affected by incompleteness in a similar
way. Similarly to the number counts, the number counts {\it ratio} of the two
subsamples is also expected to show a local maximum corresponding to stars
8 kpc away. Thus, finding the flux--color bin which maximizes the
bulge-to-disk counts ratio determines the bolometric flux of stars
at the Galactic center, and consequently \La. The power of the method stems
from the fact that the incompleteness effects nearly cancel out because a ratio
of counts is taken.

The candidate bulge stars are selected as those within a circle coinciding
with the Galactic center and radius of 10$^\circ$, except those with
$|b| < 2^\circ$ that are excluded because of confusion. The mask for the
disk star sample is defined by $|b| < 5^\circ$ and $15^\circ < |l| < 50^\circ$.

%%%%%%%%%%%%%%%%%%%%%%%%%%%%%%%%%%%%%%%%%%%%%%%%%%%%%%%%%%%%%%%%%%%%%%%
\begin{figure}
%\begin{minipage}{\textwidth}
\centering \leavevmode \psfig{file=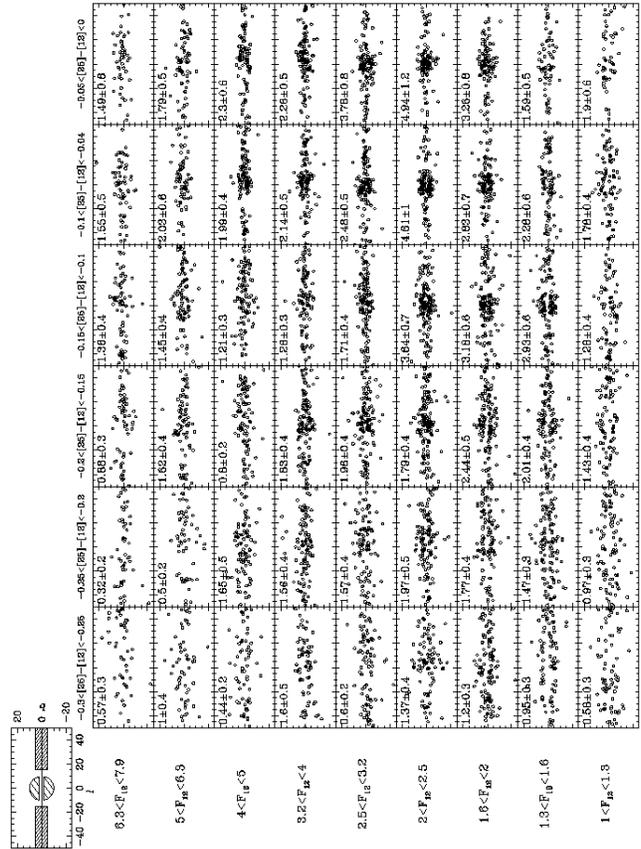,width=\hsize,clip=}
\caption{The angular distribution of stars selected by the \Fa\ flux
and [25]-[12] color. The numbers show the bulge-to-disk counts ratio
in each bin for stars selected by using angular masks shown in the
upper left corner.\label{B2Drat}}
%\end{minipage}
\end{figure}
%%%%%%%%%%%%%%%%%%%%%%%%%%%%%%%%%%%%%%%%%%%%%%%%%%%%%%%%%%%%%%%%%%%%%%%

We define 60 bins\footnote{The largest distance at which an AGB star
could be observed is strongly color dependent because the relationship
between their characteristic luminosity, \La, and observed flux, \Fa,
includes the color-dependent bolometric correction. For example,
for \La = 3,500 \Lo\ and \Fa = 1 Jy, stars with [25]-[12]= 0 can be seen
to \about 14 kpc, while stars with [25]-[12]= -0.4 only to \about 3 kpc.
This effect mandates that the analysis is performed for color ranges
sufficiently narrow that the bolometric correction is approximately constant.
Given the sample size and the behavior of the bolometric correction, we
find that 0.05 is a good choice for the color bin size.}
in the \Fa\ vs. [25]-[12] plane, and for each
determine the bulge-to-disk star count ratio (the color is limited to
$> -0.3$ because bluer stars are not detected all the way to Galactic
center, see below). Figure \ref{B2Drat} shows the angular distribution of
stars in these bins. The bulge and disk masks are shown in the upper left
corner.  It is evident that
the count ratio varies greatly among the bins, and has values indicating
both bulge detection ($> 1$) and non-detection ($\about 1$). For each
color bin there is a local maximum of the count ratio corresponding to the
stars at the Galactic center. The count ratio decreases for fainter \Fa\ flux
because stars in those bins are behind the Galactic center.

%%%%%%%%%%%%%%%%%%%%%%%%%%%%%%%%%%%%%%%%%%%%%%%%%%%%%%%%%%%%%%%%%%%%%%%
\begin{figure}
\centering \leavevmode \psfig{file=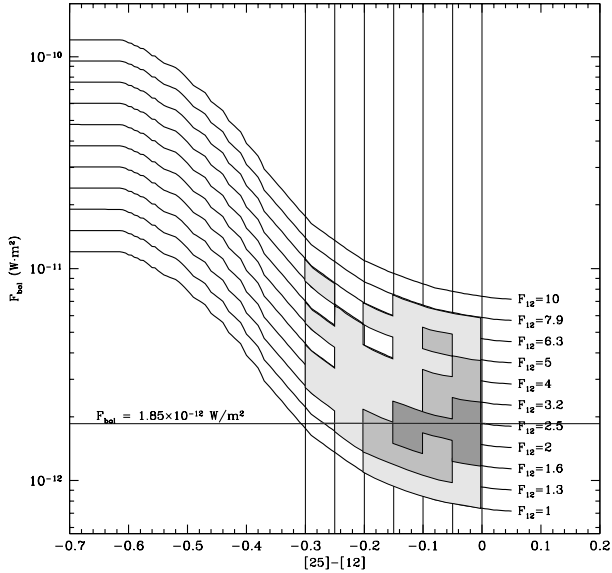,width=\hsize,clip=}
\caption{The bulge-to-disk count ratio in the \Fbol\ vs. [25]-[12] plane.
The counts ratio is shown by different shades: the darkest for $>3$, medium
for 2 to 3, and the lightest for 1 to 2.\label{shaded}}
\end{figure}
%%%%%%%%%%%%%%%%%%%%%%%%%%%%%%%%%%%%%%%%%%%%%%%%%%%%%%%%%%%%%%%%%%%%%%%

The behavior of the bulge-to-disk count ratio is easier to discern if
shown in the \Fbol\ vs. [25]-[12] plane. Figure \ref{shaded} displays
information similar to Figure \ref{B2Drat}, except now the counts ratio
is shown by different shades: the darkest for $>3$, medium for 2 to 3, and
the lightest for 1 to 2 (the bin boundaries are not rectangular
any more because \Fbol\ is used instead of \Fa). The highest contrast
is obtained for bins with [25]-[12] $\ga$ -0.2 and for \Fbol =
$1.9\times10^{12}$ Wm$^{-2}$. In principle, the highest contrast for
each color bin should be obtained at the same value of \Fbol. As
evident from the figure, the bluer bins appear to imply an \Fbol\ larger
by about a factor of 2.  However, the blue bins corresponding to \Fbol =
$1.9\times10^{12}$ Wm$^{-2}$ have \Fa\ very close to the sample completeness
limit at \Fa \about 1 Jy: so close to the sample faint limit that
even the ``counts ratio" method falls apart, and the apparent bias in
\Fbol\ for blue bins cannot be interpreted as a real effect.
We adopt \Fbol = 1.9$\times10^{12}$ Wm$^{-2}$ as the bolometric flux of
an AGB star at the Galactic center, implying \La = 3,500 \Lo\ for a distance
to the Galactic center of 8 kpc. This value could be uncertain by as much
as a factor of 2, and we estimate its probable uncertainty below.

\subsection{             Optimal Bulge Selection                      }

\begin{figure}
%\begin{minipage}{\textwidth}
\centering \leavevmode \psfig{file=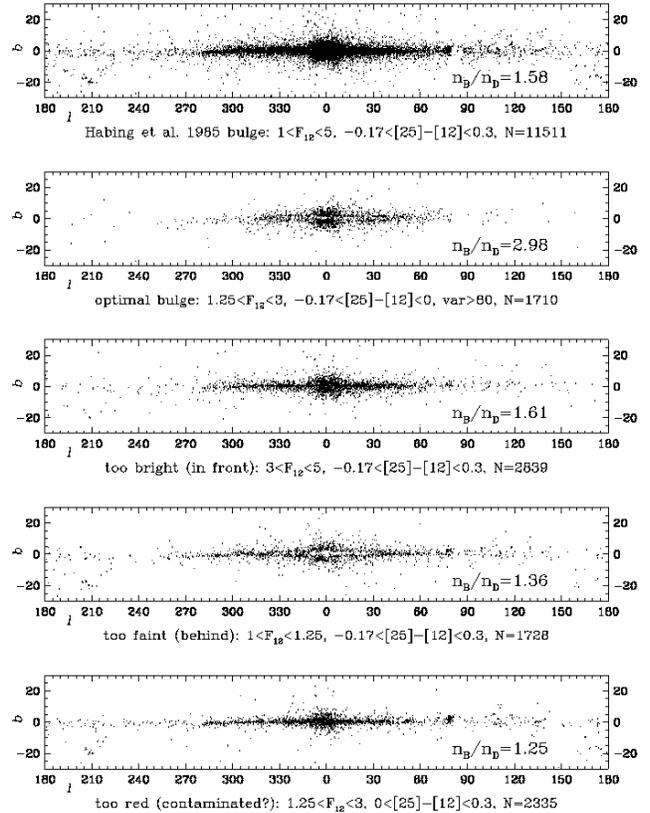,width=\hsize,clip=}
\caption{The angular distribution of stars selected by various
cuts based on the \Fa\ flux and [25]-[12] color. The numbers show
the bulge-to-disk counts ratio.\label{bulge}}
%\end{minipage}
\end{figure}

Based on the above discussion, in particular on the results presented in
Figure \ref{B2Drat}, we can optimize the bulge selection criteria to minimize
contamination. In essence, the \Fa\ vs. \Fb\ plane is mapped onto the
distance--color plane, and requiring a distance range of e.g. 7--9 kpc simply
means defining a corresponding region in the \Fa\ vs. \Fb\ plane (or,
equivalently, the \Fa\ vs. [25]-[12] plane).
We find that selecting stars with 1.25 $<$ \Fa/Jy $<$ 3.0 and $-0.17$
$<$ [25]-[12] $<$ 0.0 produces a bulge-to-disk contrast ratio of \about 3.
While this is lower than the maximal possible ratio (e.g. \about 5 for
2 $<$ \Fa/Jy $<$ 2.5 and $-0.05$ $<$ [25]-[12] $<$ 0, see Figure \ref{B2Drat}.)
the relaxed criteria produce a much larger sample (1,710 instead of 231 stars)
The bulge-to-disk contrast ratio of \about 3 is still larger than the ratio
obtained by following the prescription from Habing et al. (1985) which
produces a contrast of 1.6 (albeit with a much larger sample). The angular
distributions of stars selected by these two criteria are compared
in the two top panels in Figure \ref{bulge}.

One of the reasons why the above selection produces a larger bulge-to-disk
contrast than the Habing et al. sample is that their sample contains
disk stars which are in front of the bulge and behind the bulge, but
are observed towards the bulge. The results from Figure \ref{B2Drat}
can be used to select such stars, too. The third and the fourth panels
in Figure \ref{bulge} show the angular distribution of stars in front
of the bulge and behind the bulge, selected from the Habing et al. sample.
Another reason for the lower contrast is that their sample is probably
contaminated by non-AGB stars. This contamination would be large for
[25]-[12]$>$0, and the bottom panel in Figure \ref{bulge} shows
the angular distribution of such stars.  Their bulge-to-disk count
ratio is only 1.25.

%%%%%%%%%%%%%%%%%%%%%%%%%%%%%%%%%%%%%%%%%%%%%%%%%%%%%%%%%%%%%%%%%%%%%%%
%%%%%%%%%%%%%%%%%%%%%%%%%%%%%%%%%%%%%%%%%%%%%%%%%%%%%%%%%%%%%%%%%%%%%%%

\subsection{ Determination of the Width of the Luminosity Function  }

%%%%%%%%%%%%%%%%%%%%%%%%%%%%%%%%%%%%%%%%%%%%%%%%%%%%%%%%%%%%%%%%%%%%%%%
\begin{figure}
%\begin{minipage}{\textwidth}
\centering \leavevmode \psfig{file=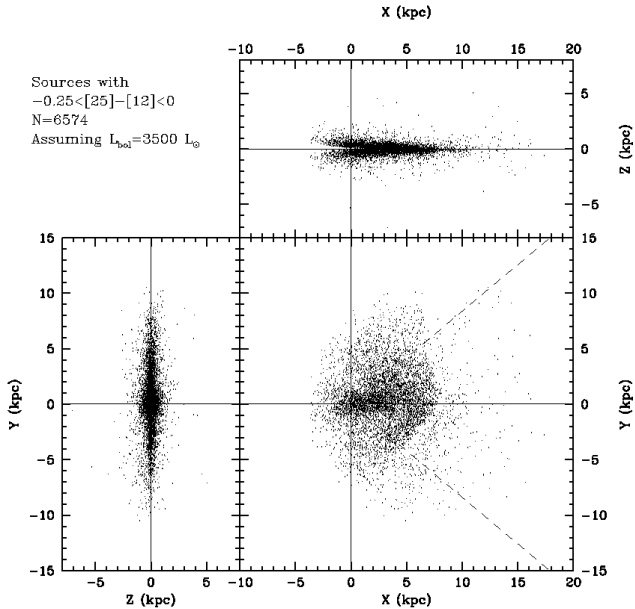,width=\hsize,clip=}
\caption{The galactic distribution of AGB stars with
-0.25 $<$ [25]-[12] $<$ 0.0, where each star is shown as a dot.
The Sun is located (X=8 kpc, Y=0, Z=0).\label{red}
}
%\end{minipage}
\end{figure}
%%%%%%%%%%%%%%%%%%%%%%%%%%%%%%%%%%%%%%%%%%%%%%%%%%%%%%%%%%%%%%%%%%%%%%%

With the assumption of a constant luminosity of 3,500 \Lo, and with the
bolometric correction shown in Figure \ref{BC}, it is straightforward to
calculate the distance for each star by using eq.~\ref{dist}. We assume
that the extinction is 0.030 mag/kpc at 12 \mic\ and 0.015 mag/kpc at 25 \mic,
resulting in corrections for stars at the Galactic center of 25\% for the
flux and $\Delta \left([25]-[12]\right) = 0.05$ for the color. In order to minimize
incompleteness effects, we require \Fa $>$ 1 Jy. This is the faintest
flux limit which still allows the detection of stars at the Galactic
center, and results in a sample of 9,926 stars. This limit also guarantees
that the sample is not affected by the IRAS faint limit at 25 \mic.
In order to visualize
the dependence of the galactic distribution of AGB stars on their [25]-[12]
color, we divide the sample into the ``blue" subsample with [25]-[12] $<$
$-0.25$ (3,352 stars), and the ``red" subsample with [25]-[12] $> -0.25$
(6,574 stars).

Figure \ref{red} shows the three Cartesian projections of the galactic
distribution of stars in the red subsample.
The X-Y and X-Z panels indicate that the sample extends beyond the Galactic center.
It is visible in the X-Z panel (upper right) that the limiting distance for
stars beyond the Galactic center depends on the height above the Galactic plane.
Stars close to the plane are observed through the bulge, and these lines of sight
have a somewhat higher faint cutoff due to the source confusion (this
effect is {\it not} caused by interstellar extinction).

The stars appear to trace out a bar-like structure of length \about 5 kpc pointing
towards the Sun, but this is an artifact of the assumption that all stars have
the same luminosity. Since the true luminosity function must have a finite width,
we can estimate this width by assuming a spherical bulge.

By analyzing the counts of stars in two strips 2 kpc wide parallel to X and Y
axes, we find that the equivalent Gaussian width
of the start count histogram along the X axis is between 1.5 and 2.5 times as
large as along the Y axis (1.2--2.0 kpc vs. 0.8 kpc). The histogram width along the X axis
is harder to measure than the width along the Y axis because it is not fully
symmetric around the X=0 due to the incompleteness effects behind the Galactic
center. We conservatively adopt 2 kpc for the effective widening of the stellar
distribution due to a finite width of the luminosity function, which implies
that the scatter (the equivalent Gaussian width) of luminosity about the mean
value is about a factor of 2. In other words, the majority of stars (66\%)
have their bolometric luminosity between 2000 \Lo\ and 7000 \Lo.
The assumption that all stars have the same luminosity smears their
distribution such that fine details cannot be recovered.

\begin{figure}
%\begin{minipage}{\textwidth}
\centering \leavevmode \psfig{file=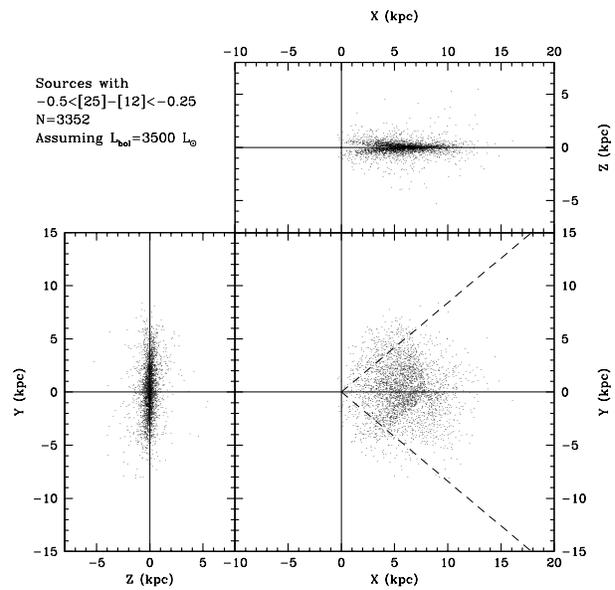,width=\hsize,clip=}
\caption{The same as previous figure, except that
-0.50 $<$ [25]-[12] $<$ -0.25.\label{blue}
}
%\end{minipage}
\end{figure}

Figure \ref{blue} shows the galactic distribution of stars in the blue subsample.
For the same bolometric flux these stars have fainter IRAS fluxes than stars
in the red subsample, and consequently a shorter limiting distance. As evident
from the figure, these stars are {\it not} observed all the way to the Galactic center.

%%%%%%%%%%%%%%%%%%%%%%%%%%%%%%%%%%%%%%%%%%%%%%%%%%%%%%%%%%%%%%%%%%%%%%%
%%%%%%%%%%%%%%%%%%%%%%%%%%%%%%%%%%%%%%%%%%%%%%%%%%%%%%%%%%%%%%%%%%%%%%%

\subsection{ Analysis of the Galactic Distribution of AGB Stars }

The Galactic distribution of AGB stars (using the luminosity function 
discussed above, \La = 3,500 \Lo) in Figures \ref{red} and \ref{blue} shows 
jumps in the number density along the lines of sight defined by $l \about 70^\circ$ 
and $l \about 320^\circ$. Both features are data artifacts. The first jump happens 
at the intersection of the IRAS missing data region with the galactic plane (the 
so-called 5$^\circ$ gap, see the IRAS Explanatory Supplement, Section III.D). 
The other jump at $l=320^\circ$ only shows up after the variability cut ($var>80$). 
This feature, as well as other more subtle structures, is a sampling effect because 
in this region the source density is high and the survey strategy produced extra 
scans at time intervals suitable for detecting variability.

In order to minimize these effects, and to study a well-defined volume,
we further limit the sample to a wedge starting at the Galactic center,
symmetric around the X axis, and with an opening angle of 80$^\circ$.
Its boundaries are shown by dashed lines in Figures \ref{red} and \ref{blue}. This
constraint leaves 6,804 stars in the sample. While the applied restriction
does not completely remove these features, it allows for a robust determination
of the source distribution inside the solar circle and several kpc beyond
it. However, around $R \about 8$ kpc any local features in the number density
should be treated with caution. These two instrumental
effects should not produce any bias in the $z$ direction, and are presumably
not dependent on color.

\subsection{          The Non-parametric Estimates}

The sample remaining after all selection cuts is still large enough
to explicitly test the hypothesis that the Galactic distribution of
AGB stars is separable in galactocentric distance, $R$, and distance
from the galactic plane, $z$. If so, then the vertical ($z$)
distributions in different radial ($R$) bins should be statistically
indistinguishable (apart from the normalization factor). We also separate
the stars by color into several subsamples in order to constrain the
relationship between the color and Galactic distribution.

\subsubsection{    The Vertical $z$ Distribution   }

\begin{figure}
%\begin{minipage}{\textwidth}
\centering \leavevmode \psfig{file=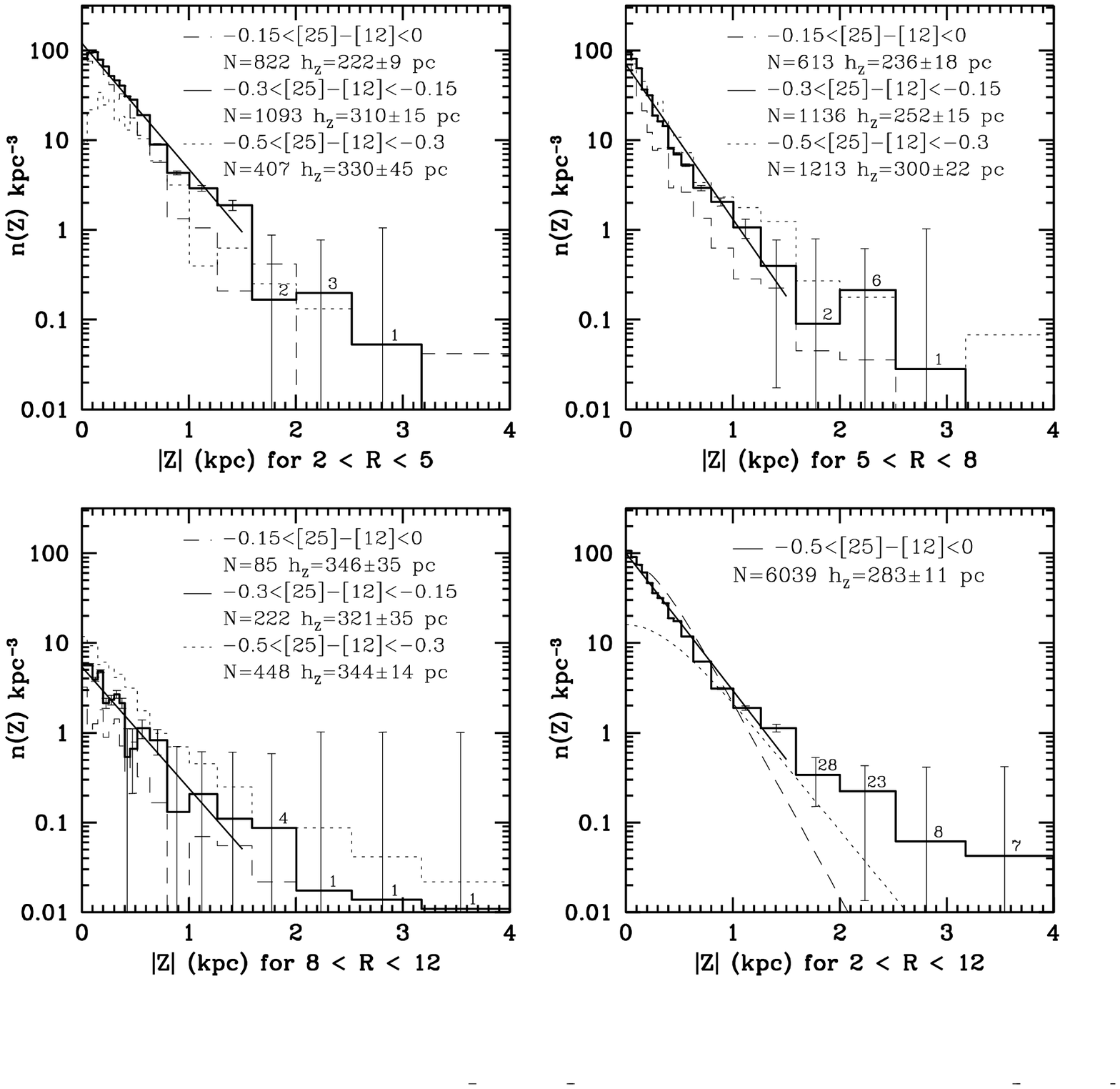,width=\hsize,clip=}
\caption{The distribution of selected AGB star candidates perpendicular
to the Galactic disk. An exponential fall-off with the scale height of
286$\pm$10 pc provides a good fit irrespective of the galactocentric
distance, $R$, and the [25]-[12] color (see text).\label{Nz}
}
%\end{minipage}
\end{figure}

Figure \ref{Nz} displays the $z$ distributions, shown as histograms
with error bars, for 3 color subsamples (the [25]-[12] color in the
ranges $-0.5$ to $-0.3$, $-0.3$ to $-0.15$, and $-0.15$ to 0), and in three
radial ranges (2-5 kpc, 5-8 kpc, and 8-12), starting with the top
left panel. The small numbers show the number of stars in each bin
(these counts are not corrected for the selection efficiency of
\about 35\%, see Sections 2.2 and 3.8). Since all histograms are well
described by an exponential function,
\eq{
        n(z) = n_o e^{{-|z| \over h_z}},
}
we determine its scale height, $h_z$, by fitting the counts for $|z|<1.5$ kpc.
The number of stars in each subsample (N) and the best-fit scale heights are shown
in each panel. The scale heights vary from 226 pc to 381 pc, with marginal evidence
(\about 3$\sigma$) that the scale height decreases with the [25]-[12] color.
While a 3$\sigma$ effect may appear
significant, we emphasize that the error bars are simply based on Poisson
statistics, and do not include any systematic effects. A similar level of
significance is obtained for the correlation between the best-fit scale
height and the radial direction, where the last radial bin appears to have
a somewhat larger scale height (as in e.g. a flared disk).

%%%%%%%%%%%%%%%%%%%%%%%%%%%%%%%%%%%%%%%%%%%%%%%%%%%%%%%%%%%%%%%%%%%%%%%
%%%%%%%%%%%%%%%%%%%%%%%%%%%%%%%%%%%%%%%%%%%%%%%%%%%%%%%%%%%%%%%%%%%%%%%

To test further whether the data strongly support these correlations,
we determine the best-fit scale height for the whole sample,
i.e. without the radial and color binning, and compare it to each
subsample. The bottom right panel in Figure \ref{Nz}  shows the $z$
distribution for the
whole sample as a histogram, and the best exponential fit by a
thin solid line. The same exponential fit (for the whole sample) is
then compared to each subsample in other three panels, and shown as
a thin solid line. We include only points with $z < 1.5$ kpc where
the signal-to-noise ratio is the highest, and use the same weight for
all points. As evident, the best-fit scale height of 286$\pm$10 pc
is not obviously inconsistent with the $z$ distribution of each subsample.
While formally the $\chi^2$ per degree of freedom is somewhat larger
than 1, the unknown systematic errors may account for this discrepancy.
We conclude that there is no compelling evidence that the scale height
depends significantly on the [25]-[12] color or the galactocentric
distance.

The histogram shown in bottom right panel of Figure \ref{Nz}
is systematically above the best-fit exponential for $|z|\ga 1.5$ kpc.
This excess of counts appears consistent with the thick disk proposed by
Gilmore and Reid (1983). However,
this discrepancy has no statistical significance; the number of stars in
all radial bins with $|z| > 1.5$ kpc is 101, or about 1.4\% of the total
number of stars, comparable to the estimated
3\% sample contamination by non-AGB stars (see Section 2.2). Such
contaminants are dominated by stars with luminosity much smaller than
the adopted AGB luminosity (3,500 \Lo). While there is growing support for
the thick disk (e.g. Chen {\em et al.} 2000, and references therein), its
existence is {\it not} required by the IRAS counts of AGB star candidates.

The dashed and dotted lines in bottom right panel of Figure \ref{Nz} show two
fits of the $sech^2$ function. This function was proposed by Habing
(1988) to be a better description of the $z$ distribution of AGB stars than
a simple exponential function, and utilized to model the IRAS counts. The
dashed line is a best-fit which reproduces the observed counts at $z \approx 0$.
However, it falls off too fast for $z \ga 1$ kpc. The dotted line produces
a good fit (to the exponential best fit) for large $z$, but it significantly
underestimates the counts for $z$ = 0. We conclude that a simple exponential
function is a much better description of the $z$ distribution of selected
AGB star candidates than $sech^2$ function for $z < 1.5$ kpc. This conclusion
is in agreement with Kent {\em et al.} (1991) who analyzed a 2.4~\mic\ map of
the northern Galactic plane.

\subsubsection{         The Radial $R$ Distribution   }

\begin{figure}
%\begin{minipage}{\textwidth}
\centering \leavevmode \psfig{file=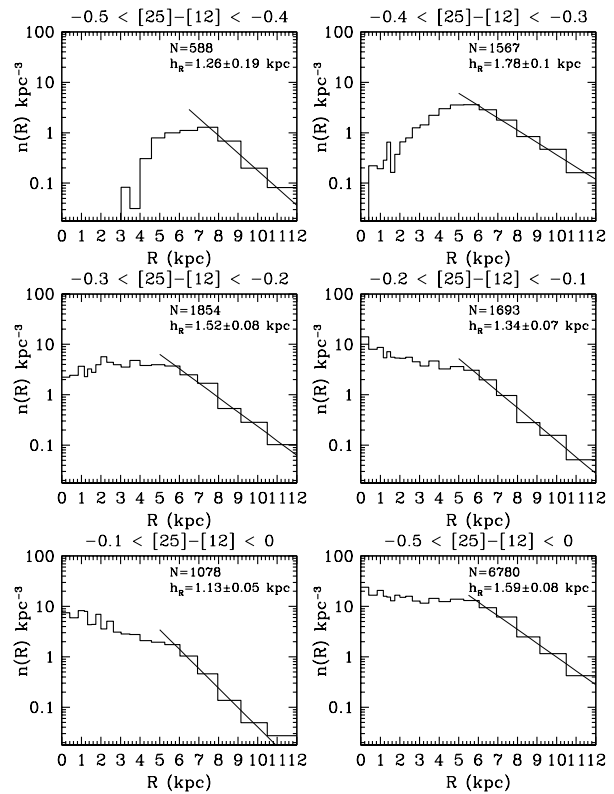,width=\hsize,clip=}
\caption{The radial dependence of AGB star number density.
\label{NR}
}
%\end{minipage}
\end{figure}

The radial distribution of selected AGB candidates is shown in Figure
\ref{NR}. The first five panels, starting in the upper left corner,
show the $R$ histograms for subsamples binned by the [25]-[12] color.
As was already evident in Figure \ref{blue}, the blue subsamples
([25]-[12] $\la$ -0.3) do not extend all the way to the Galactic center.

The decrease of number density for $R \ga$ 5 kpc seems consistent
with an exponential fall-off, and each panel displays the best-fit
scale length obtained for points with $R > 5$ kpc (except for the bluest
bin where the limit is $R > 6.5$ kpc). The corresponding fit is shown by a
straight line. The number of stars in each color-selected subsample is
also shown in each panel. Note that there are no obvious jumps in the counts
at $R$ \about 8 kpc, indicating that selecting the stars
within the wedge described in the previous section removes the
instrumental effects seen in Figures \ref{red} and \ref{blue}.

The best-fit scale lengths span the range from 1.2 kpc to 1.8 kpc,
with a typical uncertainty of 0.1 kpc, and the mean value of 1.44 kpc.
The bottom right panel shows the histogram for the whole sample
(i.e. without the color binning); its best-fit scale length is
1.6$\pm$0.07 kpc, consistent with the above mean value.

The counts of stars in the red subsamples ([25]-[12] $\ga -0.2$)
increase towards the Galactic center for $R$ $<$ 3-4 kpc. This
increase is caused by the bulge contribution. We do not attempt
to fit any analytic function because the resulting fit would
be strongly affected by the errors in the adopted luminosity
function; that is, the detailed dependence of the counts for
$R < 4-5$ kpc cannot be determined. Nevertheless, based on the
counts for the subsample with $-0.2<$[25]-[12]$< -0.1$, it appears
that the disk contribution for the inner 4$-$5 kpc may be much flatter
than its exponential fall-off inferred for larger $R$. Based on
the analysis presented in Section 3.2, the bulge contributes at
least 3 times as many stars as the disk for $R$ = 0. This implies
that the disk contribution is roughly constant within the inner
\about 5 kpc. The counts for the reddest subsample are also consistent
with this conclusion.

%%%%%%%%%%%%%%%%%%%%%%%%%%%%%%%%%%%%%%%%%%%%%%%%%%%%%%%%%%%%%%%%%%%%%
%%%%%%%%%%%%%%%%%%%%%%%%%%%%%%%%%%%%%%%%%%%%%%%%%%%%%%%%%%%%%%%%%%%%%%%

%%%%%%%%%%%%%%%%%%%%%%%%%%%%%%%%%%%%%%%%%%%%%%%%%%%%%%%%%%%%%%%%%%%%%%%
\begin{figure}
%\begin{minipage}{\textwidth}
\centering \leavevmode \psfig{file=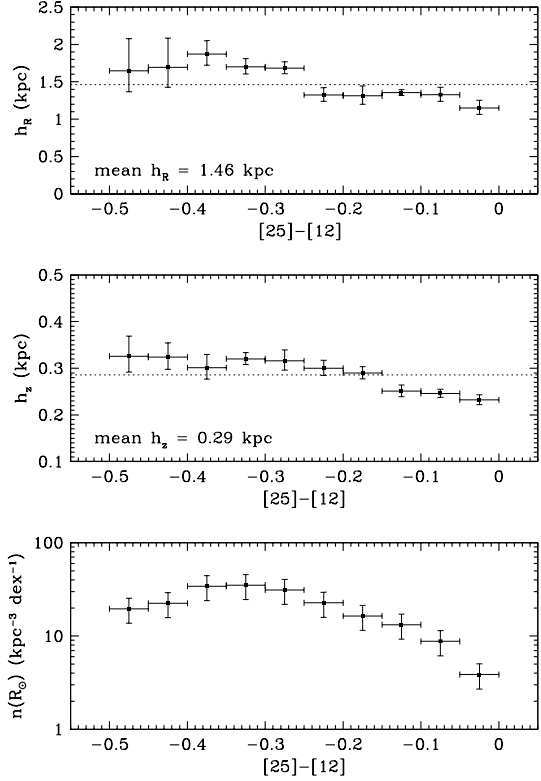,width=\hsize,clip=}
\caption{The dependence of the scale height and length on color.
\label{Ncolor}
}
%\end{minipage}
\end{figure}
%%%%%%%%%%%%%%%%%%%%%%%%%%%%%%%%%%%%%%%%%%%%%%%%%%%%%%%%%%%%%%%%%%%%%%%

\subsection{A Simple Model for the Galactic Distribution of AGB Stars}

The galactic distribution of AGB stars determined in the previous section
is only marginally inconsistent with a universal, color-independent
function which appears separable in $R$ and $z$. The $z$ distribution is
well described by an exponential function with the scale height of
\about 290 pc. The $R$ distribution has an exponential fall-off
with the scale length of \about 1.6 kpc for $R >$ 5 kpc. Within
the inner 5 kpc, the counts can be described by a flat component
due to the disk, and a bulge component which increases towards the
Galactic center.

These results do not vary strongly with the [25]-[12] color.  To
illustrate this further, the three panels in Figure \ref{Ncolor} show the
dependence of best-fit scale height, scale length and the number
density on color. This figure can be considered as a summary of results
presented in Figures \ref{Nz} and \ref{NR}. The error bars are formal
uncertainties of the fits and do not include the contributions from
the sample contamination and various incompleteness effects. Because
these contributions cannot be easily quantified, the displayed results
may be interpreted in two different ways.

Formally, it seems that both the scale height and the scale length
decrease with [25]-[12] color. Since the scale height
decreases with the stellar mass (e.g. Allen 1973), this is consistent with
the hypothesis that the mass-loss rate (which by and large controls the
[25]-[12] color, see the next section) increases with stellar mass
(see e.g. Habing 1996). The dependence of the scale length on color,
if real, implies that the high-mass stars are more concentrated towards the
Galactic center than the low-mass stars. Such a conclusion would be in agreement
with the studies of star formation inside and outside the solar circle
(Wouterloot {\em et al.} 1995, Casassus {\em et al.} 2000).

On the other hand, the statistical significance of the possible correlation
between the galactic distribution of AGB stars and their [25]-[12] color is
small. Since there are additional unknown systematic errors,
a simple universal description of the galactic distribution cannot be strongly
ruled out. Such a simple description of the distribution of AGB stars, if able
to reproduce the data, would be of great value for modeling the Galaxy, as well
as for modeling other galaxies. In order to estimate how well this model
would describe the IRAS data we perform the following test. We assume a
color-independent Galactic distribution of AGB stars as described at the beginning
of this section. The bulge is assumed to follow an exponential profile with a
scale length of 0.8 kpc (the precise form of the bulge profile is not important
since it is not well constrained by the data) with the bulge-to-disk number ratio
of 2 at $R$ = 0. Utilizing the constant luminosity \La=3,500 \Lo, we
generate the model number counts as a function of position on the sky, the \Fa\
flux and the [25]-[12] color for 100 randomly generated samples. The counts 
depend on the color despite the color-independent distribution of stars  because 
of the color-dependent
limiting distance. The discrepancy between this model and the data illustrates
the significance of the deviations from the simplified model (this test, of course,
does not reveal systematic errors, as e.g. the sample contamination).

\begin{figure}
%\begin{minipage}{\textwidth}
\centering \leavevmode \psfig{file=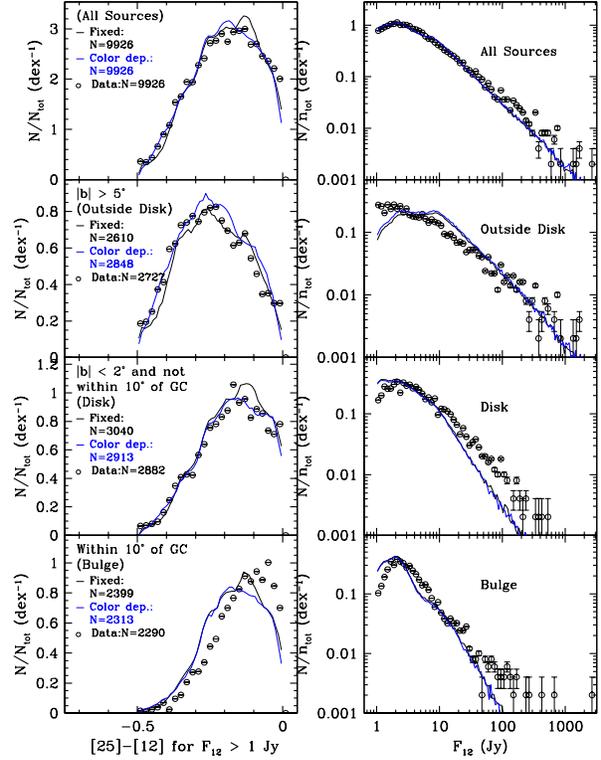,width=\hsize,clip=}
\caption{\label{model} Comparison of the IRAS data and a simplified Galactic 
model described in text.
}
%\end{minipage}
\end{figure}

Figure \ref{model} is reminiscent of Figure \ref{IRAS22xx} and shows
the comparison of the data and the simplified model. The data are shown as
squares with Poisson error bars, and the model results are shown by lines.
%For illustration, we also generated models with color-dependent scale 
%height and scale length, shown in Figure \ref{Ncolor}.
The overall normalization of the model is determined by requiring the same
total number of sources as in the data sample. The completeness function at
the faint end is determined by requiring the agreement between the data and model
counts for the whole sample (that is, the data and the model are forced to agree
in the top right panel). The real test of the model lies in the remaining panels
where the data and the model are compared for different lines of sight
{\em without} further model adjustments. The overall features in the color and flux
distributions are reproduced fairly well, although there are
some formally significant disagreements. While these disagreements illustrate
the errors introduced by applying the simplified model, they do not appear
sufficient to rule it out.

%%%%%%%%%%%%%%%%%%%%%%%%%%%%%%%%%%%%%%%%%%%%%%%%%%%%%%%%%%%%%%%%%%%%%%%
%%%%%%%%%%%%%%%%%%%%%%%%%%%%%%%%%%%%%%%%%%%%%%%%%%%%%%%%%%%%%%%%%%%%%%%

\subsection{         The number of AGB stars in the Galaxy       }

The model parameters derived in the previous section can be used to
approximately estimate the number of AGB stars in the Galaxy by direct
integration of their number density\footnote{The number of stars selected
in the 80$^\circ$ wedge, 6,591, cannot be used to directly estimate the
number of AGB stars in the Galaxy (by multiplying by 360/80) because
the blue stars are not observed all the way to the Galactic center.}.
The resulting simplified model for the number density of AGB stars with
$-0.5 <$ [25]-[12] $<$ 0 is given by
\eq{
          n(R,z) = C \, f(R) \, e^{{-|z| \over h_z}} + f_B(\sqrt{R^2+z^2}),
}
where
\eq{
                 f(R) = e^{-{R-R_c \over h_R}},
}
for $R > R_c$, and $f(R)$=1 otherwise. The radius of the inner disk part
where the number density does not depend on $R$, $R_c$, is estimated to
be 5.0 kpc. The scale length and the scale height are estimated as
$h_R=1.6$ and $h_z=0.29$ kpc, respectively. The profile of the bulge
contribution, $f_B$, is not well constrained and we assume
\eq{
             f_B(x=\sqrt{R^2+z^2}) = \zeta_B \, C \, e^{{-x \over h_B}},
}
with the characteristic length $h_B=0.8$ kpc, and the bulge-to-disk
normalization $\zeta_B = 2$.

The normalization constant $C$ can be determined from the observed
local (at the solar radius) number density of AGB stars as
\eq{
   C = {\rm exp}({\Ro-R_c \over h_R}) \,\, n(R=\Ro, z=0),
}
where the measured $n(R=\Ro,z=0)$ = 150 kpc$^{-3}$. This value is further
multiplied by a correction for the incompleteness due to selection effects
estimated to be 2.9 (the variability and color selection criteria select 35\%
of AGB stars in the adopted color range, see Section 2.2), to yield a best
estimate $C=2800$ kpc$^{-3}$. Note that the number density of AGB stars at
the Galactic center is $(1+\zeta_B)C = 3C$. It is hard to determine the associated
uncertainty which is dominated by the unknown selection effects and the
inhomogeneity of the Galactic distribution of AGB stars, but it seems that a reasonable
estimate is about a factor or 2.

The integration of the above expressions shows that the number of AGB
stars in the Galaxy is
\eq{
   N_{AGB} = C \pi \left(h_z R_c^2 + 2 h_z h_R (h_R+R_c) + 8 \zeta_B h_B^3\right).
}
Using the best estimates for the model parameters, we obtain
$N_{AGB}$ = 67.7 $C$ kpc$^3$ = 200,000. Again, this estimate is probably uncertain
to within a factor of 2.

\section{        The time evolution of the AGB mass loss       }

One of the least constrained properties of AGB stars is the time evolution
of their mass loss, and its dependence on fundamental stellar parameters,
in particular its dependence on stellar mass. Models range from a mass loss
rate that is independent of time during the AGB phase, and fully determined
by the stellar mass, to a mass loss rate that increases exponentially with time,
and is independent of the stellar mass (Habing 1996, and references
therein).

The predictions of these models relevant for the data analyzed here pertain
to the correlations between the stellar number density, color, and $z$. Since
the initial stellar mass correlates with $z$, if the mass controls the mass-loss rate,
then the [25]-[12] color, which is by and large determined by mass-loss rate,
should also correlate with $z$ (i.e. the scale height should depend on
color). At the same time, the color distribution would be a complicate
convolution of the initial mass function and the evolutionary time scales
(which also depend on the stellar mass). On the other hand, if the
mass-loss rate does not depend on the stellar mass, then there should be no
correlation between the scale height and color, and the distribution of
colors would reflect the temporal evolution of mass-loss rate.

As shown in the middle panel in
Figure \ref{Ncolor}, there is some evidence that the scale height
decreases with the [25]-[12] color, as would be the case if only
the high-mass stars develop large mass-loss rates. Since this
result may be caused by various systematic effects, we only explore
the alternative possibility that the mass-loss rate does not depend on
the stellar mass, and is a universal function of time. This approach
assumes that the scale height and scale length do not depend on the
[25]-[12] color, and interprets the variation of the number density
with color as due to temporal evolution of the mass-loss rate (see
the bottom panel in Figure \ref{Ncolor}). While in principle both
the variation of the scale height and the number density with color
could be used to simultaneously constrain the dependence of mass-loss
rate on stellar mass {\it and} time, the available data are not sufficient
to derive robust conclusions in such a two-dimensional problem.

\begin{figure}
%\begin{minipage}{\textwidth}
\centering \leavevmode \psfig{file=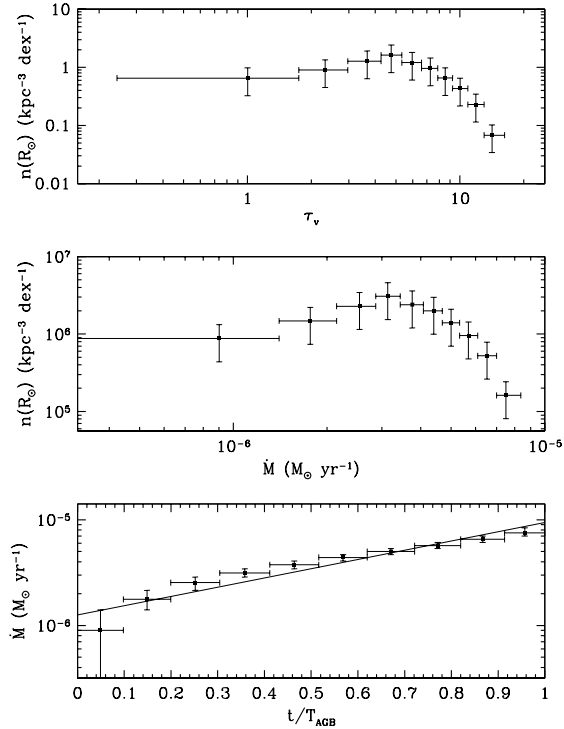,width=\hsize,clip=}
\caption{\label{Mdot} Derived AGB mass-loss rate as a function of $t/T_{AGB}$, 
where $T_{AGB}$ is the time spent in the observed phase of AGB stellar evolution. 
}
%\end{minipage}
\end{figure}

For given dust grains, the [25]-[12] color is essentially fully determined
by the dust optical depth (IE95, IE97). We use a model derived relationship
(the same models are used to derive the bolometric correction discussed in
Section 3.1) between the [25]-[12] color and the visual optical depth, $\tau_V$,
to transform the number density vs. color relation shown in the bottom panel
of Figure \ref{Ncolor}, to the number density vs. $\tau_V$ relation shown in the
top panel of Figure \ref{Mdot}.

For given luminosity, dust grains, and dust-to-gas ratio, the dust optical
depth is by and large determined by the mass-loss rate (Bedijn 1987, IE95).
We use a relationship derived from radiatively driven wind models with
silicate dust by Elitzur \& Ivezi\'c (2001). Assuming a standard gas-to-dust
mass ratio (200),
\eq{
    \Mdot = 0.9 \times 10^{-6} \, \tau_V^{3/4} \, \Mo \, {\rm yr}^{-1}
}
Note that \Mdot\ $\propto \tau_V^{3/4}$ rather than \Mdot\ $\propto \tau_V$
due to dust drift effects. The resulting number density vs. mass-loss rate relation
is shown in the middle panel of Figure \ref{Mdot}. The observed mass-loss rate
spans the range from \about 10$^{-6}$ \Mo/yr to \about 10$^{-5}$ \Mo/yr (this
particular range is a consequence of the analyzed range of [25]-[12] color).

The temporal behavior of the mass-loss rate is directly reflected in
the observed distribution of mass-loss rate, assuming that the same
function applies to all stars in the sample. For example, if the mass-loss
rate increases quickly with time between two values, then most stars
will be observed with a mass-loss rate closer to the low value. Following
this assumption, we derive the mass-loss rate distribution shown by
symbols in the bottom panel of Figure \ref{Mdot}. The error bars are computed
by assuming Poisson statistics. As evident, the increase of mass-loss rate is
well described by an exponential function (i.e. log(\Mdot) $\propto t$), and
a best fit is displayed by the line.

Note that the bottom panel of Figure \ref{Mdot} displays the mass-loss rate
as a function of $t/T_{AGB}$, where $T_{AGB}$ is the time spent in the observed
phase of AGB stellar evolution (not necessarily equal to the entire duration of
the AGB phase due to the applied color limits). This time cannot be determined
from the data analyzed here, and the only constraint we have on the mass-loss rate
temporal behavior is that it is well described by an exponential function.
The total number of AGB stars in the Galaxy (estimated here to be about 200,000)
can be used to estimate $T_{AGB}$ only if the number of stars in the Galaxy that
become AGB stars is known, as well as their mean lifetime. Assuming $2\times10^{10}$
for the former, and $10^{10}$ years for the latter, reproduces the canonical AGB
lifetime of $10^5$ years. Increasing the number of stars that undergo the AGB
phase decreases the estimate of $T_{AGB}$.

%%%%%%%%%%%%%%%%%%%%%%%%%%%%%%%%%%%%%%%%%%%%%%%%%%%%%%%%%%%%%%%%%%%%%%%
%%%%%%%%%%%%%%%%%%%%%%%%%%%%%%%%%%%%%%%%%%%%%%%%%%%%%%%%%%%%%%%%%%%%%%%

\section{                        Discussion                       }

\subsection{  Summary of the Results}

We assume that AGB stars can be reliably selected by using the
IRAS variability index and the [25]-[12] color, that their
bolometric flux can be estimated from the IRAS \Fa\ flux and the
[25]-[12] color, and that they all have the same luminosity.
We estimate this luminosity to be 3,500 \Lo, and determine
the distance to each star. The analysis of the resulting
galactic distribution shows that it is well described by a
simple function independent of color.

This is the first direct (i.e. not based on fitting the number
counts) estimate of the galactic distribution of AGB stars throughout
the Galaxy. It is somewhat surprising that a good description
of the IRAS observations can be obtained by a model including only five
free parameters: the scale height (\about 290 pc), the radius of the
inner disk where the number density does not depend on the galactocentric
radius (\about 5 kpc), the scale length (\about 1.6 kpc) for the exponential
fall-off in the outer disk, the bulge-to-disk number density ratio at the
Galactic center (the bulge density is twice as large as the disk density),
and the overall normalization (the local density in the disk plane at
the solar radius of \about 100 AGB stars per kpc$^{-3}$ for stars with
$-0.5 <$ [25]-[12] $<$ 0). The normalization is probably accurate
to better than a factor of 2, and other parameters to within about
20\%. The simplified model presented here implies that there are about
10$^5$ AGB stars in the Galaxy, with the uncertainty of about a factor
of 2. While there exist other models that provide excellent description
of IRAS counts (e.g. Wainscoat {\em et al.} 1992), they are usually much
more involved that the model presented here, and often are not uniquely
determined.

The final simplified model is very similar to models derived in several
other studies, which are based on different methods for estimating distance. For example,
Jura \& Kleinmann (1992) studied the vertical scale height for \about 300 Mira stars
with $|b| > 30^\circ$ whose distance was determined from the period-luminosity
relation. They determine the scale height of \about 240 pc for stars
with periods, $P$, longer than 300 days, in agreement with the results derived
here (the period and color are correlated such that redder stars have
longer periods, see Habing 1996). For stars with period shorter than 300
days they found a scale height of 500-600 pc. We do not find evidence
for such an increase of the scale height for the blue subsamples. Either
our selection procedure missed a significant fraction of the blue stars
close to the faint end, or some stars in the sample discussed by Jura \&
Kleinmann had overestimated distance. Indeed, they point out that
the stars in their sample with $P <$ 300 days have about twice as large a
velocity dispersion in the $z$ direction (55 km/s) as stars with $P>$ 300 day.
This may indicate that the former are contaminated by non-AGB stars.

Blommaert, van der Veen \& Habing (1993) found that the number density of
(red) AGB stars falls off more steeply outside the solar circle than
in the inner Galaxy. This finding agrees well with the change of slope at
\about 5 kpc evident in Figure \ref{NR}. The radial scale length determined
here (1.6 kpc) is at the low end of estimates in the literature (1.8--6 kpc,
Kent, Dame \& Fazio 1991, and references therein). However, the direct 
comparison is inappropriate because our value is determined for $R>5$ kpc,
while in most other studies the assumed exponential profile extends to
the Galactic center. Fitting this function to data shown in the bottom 
right panel in Figure \ref{NR}, we obtain a scale length of 3$\pm$1 kpc,
in agreement with recent studies. It is noteworthy that studies based
on infrared data yield systematically smaller scale lengths than optical
studies (Wainscoat {\em et al.} 1992, and references therein).

Jura, Yamamoto \& Kleinmann (1993) found that the number ratio of
stars with P $>$ 400 days and stars with 300 days $<$ P $<$ 400 days is
larger at \about 1 kpc from the Galactic center (\about 0.7) than locally
(only \about 1/6). This implies that the counts ratio of blue to
red AGB stars should be lower close to the Galactic center. However,
the analysis presented here indicates that the counts of blue
stars drop for R $<$ 2 kpc due to the IRAS flux limit.
That is, we find no evidence that the ratio of red to blue AGB stars
varies across the Galaxy. While this ratio may as well be different in
the bulge, this is not required by the IRAS data.

\subsection{         Pitfalls   }

Of course, all of the above results critically depend on the various
adopted assumptions. For example, although the estimated sample contamination
is very low (\about 3\%, see section 2.2), it could be somewhat higher
because the AGB nature of these stars is not positively determined for each
star. Similarly, the selection does not appear biased with respect to the
\Fa\ flux and the [25]-[12] color, but this conclusion is also based on
statistical arguments. The employed model-derived bolometric correction
implies that the SEDs of all AGB stars with silicate dust are self-similar.
While this assumption is certainly not strictly true, it appears to be
correct to within a factor of \about 2. A systematic bias with respect to
color of the bolometric correction could produce false evidence for the
dependence of the derived scale height and scale length on color. Some
evidence for such dependence is borne by the data (see Sections 3.6 and
3.7, and Figure \ref{Ncolor}), but because of the  bolometric correction
uncertainties, it is not clear whether this effect is real.

It is obvious that the true AGB star luminosity function is not a
$\delta$-function; yet the shape of the distribution of stars
around the Galactic center implies that the majority of stars have
luminosity within a factor of 2 from the median value. The assumption
that all stars have the same luminosity results in smearing of
their distribution such that fine details cannot be recovered.
This effect may hide some interesting features, but it does
not strongly affect the overall stellar distribution.

Due to all these uncertainties, the temporal behaviour of mass loss on
the AGB cannot be strongly constrained. If the dependence of the scale
height and scale length is real, then the mass-loss rate increases with
the stellar mass. Alternatively, if this dependence is dismissed as
due to systematic effects, then the observed color distribution
implies that the mass-loss rate increases exponentially with
time.

\subsection{   Possibilities for Improvement    }

This work demonstrates that infrared observations of AGB stars are an
excellent tool for studying the Galactic structure all the way
to its center and beyond. It also reveals all the pitfalls
associated with the limited data set. Fortunately, these shortcomings
are solvable in principle, and the AGB stars could be utilized in
a study with much greater statistical power than possible with only
the IRAS data.

The reliability of AGB star selection could be improved by multi-wavelength
multi-epoch observations, e.g. as those obtained by van der Veen \& Habing
(1990), or Whitelock, Feast \& Catchpole (1991). Because of the characteristic
shape of SED, and its variability properties, such observations can be
used to reliably separate AGB stars from other similar sources. Additionally,
the observations of various masers  could be utilized as yet another
signature of AGB phase (as in e.g. Jiang {\em et al.} 1991). A further important
gain from the multi-wavelength observations is the ability to determine the
bolometric flux directly, rather than by using a bolometric correction.

A study of the Galactic distribution of AGB stars would greatly benefit
from an all-sky survey about 10 times more sensitive than IRAS. Such a survey would
be capable of detecting AGB stars of all colors beyond the Galactic center,
rather than only those with e.g. [25]-[12] $\ga -0.3$, as with the IRAS data.
By utilizing the fact that the distribution of stars is symmetric around the
Galactic center, the hypothesis that the characteristic AGB star luminosity
does not depend on the [25]-[12] color could be explicitly tested on a large
sample of stars. Even if obtained only for the 10$^\circ \times 10^\circ$ area
toward the Galactic center, such a survey would provide significant new constraints
for the evolution of AGB stars and their Galactic distribution. We are currently
investigating the possibility of using 2MASS and SIRTF surveys for such a
study.

\section*{Acknowledgments}

We acknowledge generous support by Princeton University, and by NASA grants 
NAG5-6734 and NAG5-11094 to GRK. We thank Tom Chester for clarifying the definition 
of IRAS variability index.

\label{lastpage}
\end{document}